\documentclass[twocolumn,a4paper,reprint,amssymb,aps,prd,groupedaddress,superscriptaddress,nofootinbib]{revtex4-2}
\pdfoutput=1
\usepackage{graphicx}
\usepackage{epsf}
\usepackage{bm}
\usepackage{amsmath}
\usepackage{amsfonts}
\usepackage{amssymb}
\usepackage{epstopdf}
\usepackage{natbib}
\usepackage{color}
\usepackage[dvipsnames]{xcolor}
\usepackage{physics}
\usepackage[colorlinks = true,
            linkcolor = blue,
            urlcolor  = blue,
            citecolor = blue,
            anchorcolor = blue]{hyperref}
\providecommand{\U}[1]{\protect\rule{.1in}{.1in}}
\usepackage[capitalize]{cleveref}

\usepackage{soul}

\newcommand{\be}{\begin{equation}}
\newcommand{\ee}{\end{equation}}
\newcommand{\bea}{\begin{eqnarray}}
\newcommand{\eea}{\end{eqnarray}}

\begin{document}

\title{Is the dark energy equation of state parameter singular?}

\author{Emre \"{O}z\"{u}lker}
\email{ozulker17@itu.edu.tr}
\affiliation{Department of Physics, Istanbul Technical University, Maslak 34469 Istanbul, Turkey}

\begin{abstract}
A dark energy with a negative energy density in the past can simultaneously address various cosmological tensions, and if it is to be positive today to drive the observed acceleration of the universe, we show that it should have a pole in its equation of state parameter. More precisely, in a spatially uniform universe, a perfect fluid (submitting to the usual continuity equation of local energy conservation) whose energy density $\rho(z)$
vanishes at an isolated zero $z=z_p$, necessarily has a pole in its equation of state parameter $w(z)$ at $z_p$, and, $w(z)$ diverges to positive infinity in the limit $z\to z_p^+$ and it diverges to negative infinity in the limit $z\to z_p^-$---we assume that $z_p$ is not an accumulation point for poles of $w(z)$.
 However, the converse statement that this kind of a pole of $w(z)$ corresponds to a vanishing energy density at that point is not true as we show by a counterexample. An immediate implication of this result is that one should be hesitant to observationally reconstruct the equation of state parameter of the dark energy directly, and rather infer it from a directly reconstructed dark energy density.
\end{abstract}

\maketitle

\section{Introduction}
The $\Lambda$CDM model is the standard model of cosmology based on the theory of general relativity with a cosmological constant ($\Lambda$) and the existence of a pressureless source dubbed cold dark matter (CDM). Despite its immense success and reign of over two decades, its persistent inefficacy in explaining certain cosmological observations advocates the need for an update to the standard model of cosmology~\cite{DiValentino:2020zio,DiValentino:2020vvd,DiValentino:2020srs,DiValentino:2021izs,Perivolaropoulos:2021jda,Abdalla:2022yfr,DiValentino:2020vhf}.

One way of tackling the shortcomings of $\Lambda$CDM is by considering cosmological models incorporating, on top of the usual sources of $\Lambda$CDM (matter, radiation, etc.), a dark energy (DE) perfect fluid (as an effective or actual source) with energy density $\rho_{\rm DE}(z)$ that attains negative values in the past. Out of the various observational tensions arising within $\Lambda$CDM that can be addressed via a negative DE density, the most prominent ones are the Hubble constant ($H_0$), Lyman-$\alpha$ (Ly-$\alpha$) and $S_8$ discrepancies. The $S_8$ discrepancy is indirectly, the former two discrepancies are directly related to the value of
\begin{equation}
    3H^2(z)=\rho_{\rm m0}(1+z)^3+\rho_{\rm DE}(z),\label{eq:friedmann}
\end{equation}
where we work in in units $c=1$ and $8\pi G=1$, $z$ and $H(z)$ are respectively the redshift and the Hubble function, $\rho_{\rm m0}$ is the present-day matter density, $\rho_{\rm DE}(z)$ is the DE density (the cosmological constant for $\Lambda$CDM), and radiation, i.e., photons and relativistic relics, are neglected. First, the local estimations of the Hubble constant---e.g., ${H_{0}=73.04\pm1.04~{\rm km\, s^{-1}\, Mpc^{-1}}}$ from the calibration of Supernovae using Cepheid variables \cite{Riess:2021jrx} and ${H_{0}=69.8\pm0.8~{\rm km\, s^{-1}\, Mpc^{-1}}}$ when they are calibrated via the Tip of the Red Giant Branch (TRGB) \cite{Freedman:2019jwv}---are in tension at various levels of significance with its inferred values from observational constraints when cosmic microwave background (CMB) data is included in the analyses; e.g., \textit{Planck} Collaboration found $H_0=67.36\pm0.54~{\rm km\, s^{-1}\, Mpc^{-1}}$ from the full CMB data \cite{Planck:2018vyg}. Second, while the observations prefer a higher value of the Hubble constant compared to \textit{Planck} results for $\Lambda$CDM, even the most recent Ly-$\alpha$ measurements of eBOSS (SDSS DR16) still prefer a lower $H(z)$ value at the effective redshift $z=2.33$, and also a higher comoving angular diameter distance to this redshift which is also achievable by lower values of ${H(z)}$ for $z<2.33$, showing a mild $\sim1.5\sigma$ tension with the \textit{Planck} results \cite{duMasdesBourboux:2020pck,eBOSS:2020yzd}. Third, the $S_8\equiv\sigma_8\sqrt{\Omega_{\rm m0}/0.3}$ parameter is used to quantify the discrepancy in the $\sigma_8-\Omega_{\rm m0}$ plane where $\sigma_8$ is a measure of the amount of structure defined as the root-mean-square of the present-day matter density fluctuations within spheres of $8h^{-1}\,\rm Mpc$, ${h\equiv H_0/100{\rm km\, s^{-1}\, Mpc^{-1}}}$ is the dimensionless reduced Hubble constant and $\Omega_{\rm m0}\equiv\rho_{\rm m0}/3H_0^2$ is the present-day matter density parameter. Similar to the previous discrepancies, the measurements of $S_8$ from low redshift probes (weak lensing, cluster counts, redshift-space distortion) do not agree well with its inferred value when $\Lambda$CDM is analyzed using CMB; e.g., compare the \textit{Planck} result of $S_8=0.832\pm0.013$ from full CMB data \cite{Planck:2018vyg} with $S_8=0.766_{-0.014}^{+0.020}$ from KiDS-1000 (weak lensing) \cite{KiDS:2020suj}. $S_8$ is a model-dependent parameter that is sensitive to the expansion history, $H(z)$, both through $\sigma_8$ and $\Omega_{\rm m0}$---the effect of $H(z)$ on $\Omega_{\rm m0}$ is clear from the definition of the present-day density parameter, and growth of structure is naturally affected by the speed and duration of the expansion that work against it on large scales. Considering the nature of the above discrepancies, a particular $H(z)$ function may solve all three of them, and, as seen from \cref{eq:friedmann}, this $H(z)$ can be translated to a particular form of the effective DE density $\rho_{\rm DE}(z)$ when $\rho_{\rm m0}$ is constrained. Nonparametric observational reconstructions of $H(z)$ or $\rho_{\rm DE}(z)$ aiming to get hold of such functions from these data, consistently find $\rho_{\rm DE}(z)$ to attain negative values (although usually statistically consistent with zero) for $z\gtrsim1.5$ \cite{Bonilla:2020wbn,Sahni:2014ooa,Aubourg:2014yra,Wang:2018fng,Poulin:2018zxs,Escamilla:2021uoj}. Moreover, parametric reconstructions and models that allow the DE density to attain negative values, can relax some or all of the $H_0$, $S_8$ and Ly-$\alpha$ discrepancies by invoking a negative DE density at $z\gtrsim1.5$ \cite{Akarsu:2023,Visinelli:2019qqu,Sen:2021wld,Calderon:2020hoc,Sahni:2014ooa,DiValentino:2020naf,Akarsu:2019hmw,Dutta:2018vmq,Akarsu:2021fol,Akarsu:2019ygx,Acquaviva:2021jov,Akarsu:2022lhx}. In particular, in Ref.~\cite{Akarsu:2019hmw}, the graduated DE (gDE) was shown to ameliorate the $H_0$ and Ly-$\alpha$ discrepancies by preferring a density that rapidly transitions from negative to positive, and in Ref.~\cite{Akarsu:2021fol}, its limiting case the sign switching cosmological constant ($\Lambda_{\rm s}$CDM) where the transition happens instantaneously, was shown to ameliorate all three discrepancies; the upcoming observational analysis of $\Lambda_{\rm s}$CDM with an extended dataset that includes a prior on the absolute luminosity magnitudes of Supernovae, shows strong preference of the model over $\Lambda$CDM in Bayesian evidence \cite{Akarsu:2023}. All of the above points show strong phenomenological motivation for considering a DE density that attains negative values in the past.

A negative energy density may seem alarming from a theoretical perspective with regards to the established energy conditions, which are usually considered along with general relativity (GR), viz., the dominant energy condition (DEC), the weak energy condition (WEC) etc. We start addressing these concerns by first noting that while a perfect fluid (effective) DE density that is negative definitely violates the DEC and WEC---it also violates the null energy condition (NEC) and the strong energy condition (SEC) if it continuously transits to a positive value in the late universe, due to the behavior studied in this work of its equation of state (EoS) parameter $w_{\rm DE}$---this does not necessarily mean it is incompatible with the singularity/censorship theorems that are proved relying on these energy conditions; that is because, the singularity theorems rely on restrictions of the Einstein tensor corresponding to the restrictions imposed on the total energy momentum tensor (might be effective) by the energy conditions, and, in the presence of multiple sources or modified gravity, the total energy momentum tensor may submit to these energy conditions while the DE itself does not---an example is the nonminimally interacting DE (IDE) models~\cite{Salvatelli:2014zta} for which the DE density can become arbitrarily negative as long as it is compensated by the rest of the sources whose energy density contributing to the total energy momentum tensor is positive. Second, the energy conditions are reasonable assumptions that any field obeys, but, there is no \textit{a priori} reason that they should hold~\cite{Curiel:2014zba}. In fact, all pointwise energy conditions (which include all the examples given above) are violated by quantum fields~\cite{Epstein:1965zza}; even one of the weakest nonpointwise energy conditions, the so called averaged null energy condition (ANEC) does not always hold for quantum fields~\cite{Wald:1991xn}. Perhaps more suspiciously, simplest classical configurations with a scalar field such as quintessence generically violates the SEC, and a nonminimally coupled scalar field as in Brans-Dicke theory~\cite{Brans:1961sx} can violate all the pointwise energy conditions and even the ANEC~\cite{Visser:1999de}.
Third, the DE need not be a physical energy source, but rather an effective term in the field equations due to an underlying modified theory of gravity (some examples are provided below) in which case whether it satisfies the energy conditions or not is a matter of interest but not an expectation.

Considering the above points, and that there exists a plethora of studies in the literature on phantom (${w_{\rm DE}<-1}$) DE models all of which violate all the pointwise energy conditions similar to a negative DE density, we encourage not being reluctant with regards to the possible negativity of the DE. Theoretically motivated models that incorporate an actual or effective negative energy source are already present in the cosmology literature, e.g., ghost-matter cosmologies~\cite{Chavda:2020tfh}, scalar-tensor theories of gravity such as Brans-Dicke theory~\cite{Faraoni:1998qx}, loop quantum gravity~\cite{Ashtekar:2011ni}, braneworld models~\cite{Sahni:2002dx}, everpresent $\Lambda$~\cite{Ahmed:2002mj}, unimodular gravity without energy conservation~\cite{Perez:2020cwa}, quadratic bimetric gravity~\cite{Mortsell:2018mfj}, $f(T)$ gravity~\cite{Alvarenga:2012bt}, $f(T^{\mu\nu}T_{\mu\nu})$ gravity (also known as energy-momentum squared gravity)~\cite{Akarsu:2019ygx}, Snyder-de Sitter scalar~\cite{Franchino-Vinas:2021bcl}, a dynamical cosmological term $\Lambda(t)$~\cite{Grande:2006nn}; see also Ref.~\cite{Akarsu:2021fol} and references therein. Also, recently an anti-de Sitter to de Sitter transition in line with the conjecture of Ref.~\cite{Akarsu:2019hmw} was suggested relying on running Barrow entropy~\cite{DiGennaro:2022ykp}.

Of course such a negative DE density should transit to positive regime at late times to drive the present-day acceleration of the universe, and if it varies continuously, it should vanish at least for once at a redshift $z=z_p$. It has been observed that such models present a pole in the DE equation of state (EoS) parameter $w_{\rm DE}(z)\equiv p_{\rm DE}(z)/\rho_{\rm DE}(z)$ at the transition point $z_p$ ($p_{\rm DE}$ denotes the DE pressure); moreover, this pole results in an EoS parameter that diverges to negative infinity if we approach to pole in the direction $z\to z_p^-$, and diverges to positive infinity in the direction $z\to z_p^+$~\cite{Sahni:2014ooa,Wang:2018fng,DiValentino:2020naf,Akarsu:2019hmw,Escamilla:2021uoj,Gomez-Valent:2015pia,Sahni:2004fb,Tsujikawa:2008uc,Zhou:2009cy,Bauer:2010wj,Sahni:2002dx,Akarsu:2019ygx,Acquaviva:2021jov,Akarsu:2022lhx}.
 See the bottom panel of Fig.~\ref{fig:ex} for this type of divergence and note that, for an expanding spatially uniform universe, the limit $z\to z_p^+$ corresponds to moving forward in time while $z\to z_p^-$ to backwards. A particular cosmological scenario in which the EoS parameter becomes singular as the total energy density of the universe vanishes was first investigated in Ref.~\cite{Dabrowski:2009kg}, and it was shown in Ref.~\cite{Sahni:2014ooa} that for any effective DE whose energy density vanishes at $z_p$, the effective EoS parameter diverges such that $\abs{w_{\rm DE}(z_p)}\to\infty$, provided that $z_p$ is not a critical point of $\rho(z)$ and any other sources accompanying the DE are pressureless. In this paper, we prove that this kind of a pole is a necessary condition for any source submitting to the local energy conservation if it attains a vanishing energy density even at a critical point regardless of the accompanying sources, moreover, the divergence of $w(z)$ has the above mentioned characteristic with opposite signs around the pole as depicted in Fig.~\ref{fig:ex}, and also that the converse is not necessarily true, i.e., this kind of a pole does not imply that the DE density vanishes at that point.

\section{statement}
Let us consider a perfect fluid source (possibly multicomponent) described by the energy momentum tensor (EMT) $T_{\mu\nu}=(\rho+p) u_\mu u_\nu+p g_{\mu\nu}$, where $g_{\mu\nu}$ is the metric, $\rho$ and $p$ are respectively the energy density and pressure of the fluid as measured in its rest frame, $u^\mu$ is the timelike vector field describing the 4-velocity of the fluid, and we work in units $c=1$ and $8\pi G=1$ with the metric signature $(-\,+\,+\,+)$ resulting in $u^\mu u_\mu=-1$. Einstein field equations (EFE) read
\begin{equation}
    G_{\mu\nu}=T_{\mu\nu},
    \label{eq:EFE}
\end{equation}
where $G_{\mu\nu}$ is the Einstein tensor, and through the twice contracted Bianchi identity, leading to $\nabla^\mu G_{\mu\nu}=0$, they imply the local energy conservation 
\be
\nabla^\mu T_{\mu\nu}=0,
\label{eq:localcons}
\ee
where $\nabla$ is the metric-compatible covariant derivative. From now on, in accordance with the cosmological principle, we assume a universe described by the Robertson-Walker (RW) metric $\dd{s}^2=-\dd{t}^2+a^2(t)\dd{\Sigma}^2,$
where $t$ is cosmic time, $a(t)$ is the scale factor with $a_0$ being its present day value, $\dd{\Sigma}$ is the metric of three dimensional homogeneous and isotropic spatial hypersurfaces, and we define the redshift as $z\equiv-1+ a_0/a$. Given the RW metric,  \cref{eq:localcons} implies the continuity equation
\be
\dv{\rho}{t}+3H(\rho+p)=0,\label{eq:cont}
\ee
where $H\equiv\frac{1}{a}\dv{a}{t}$ is the Hubble parameter.

\begin{figure}
    \centering
    \includegraphics[width=0.45\textwidth]{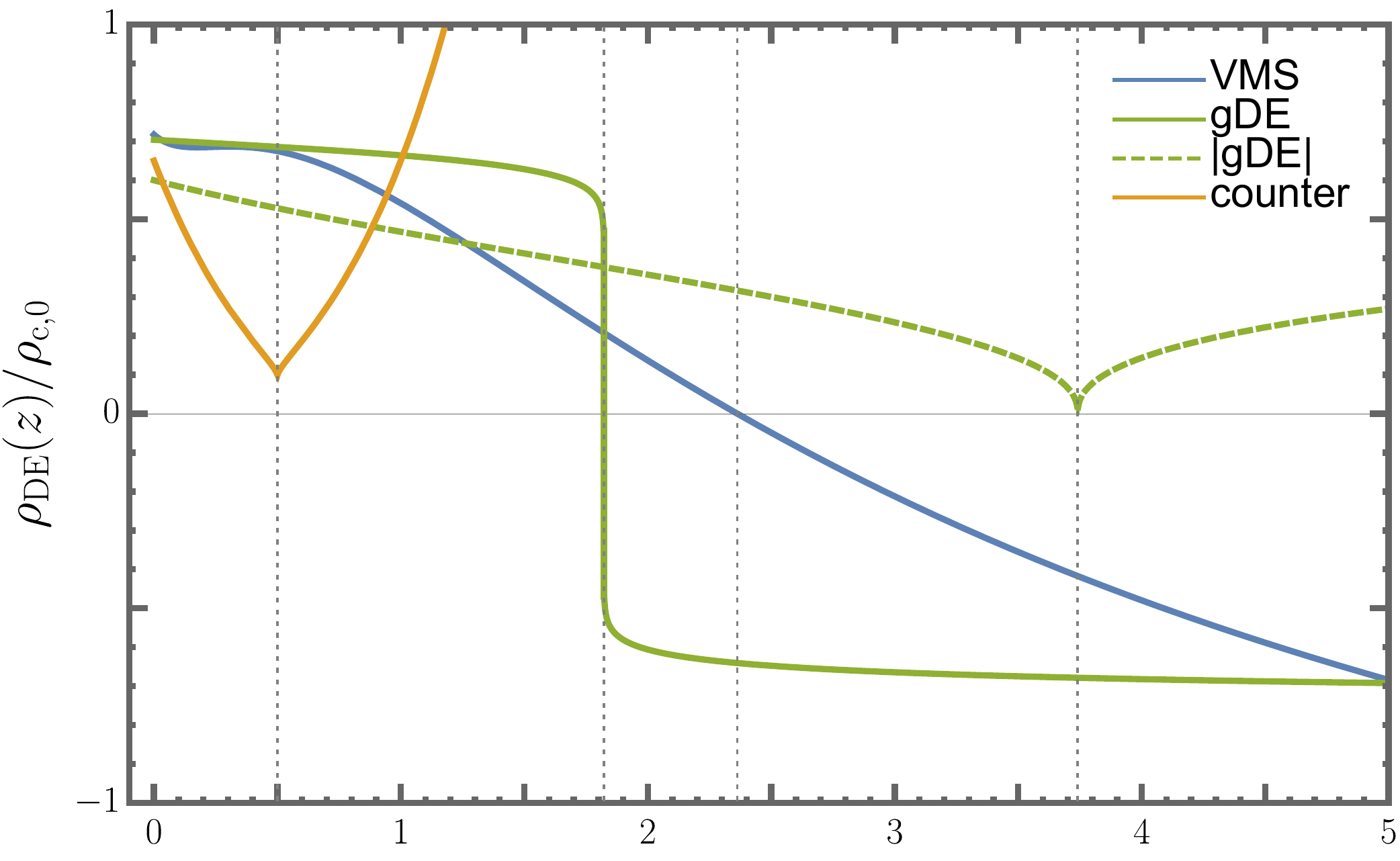}
    \hspace{1cm}\includegraphics[width=0.45\textwidth]{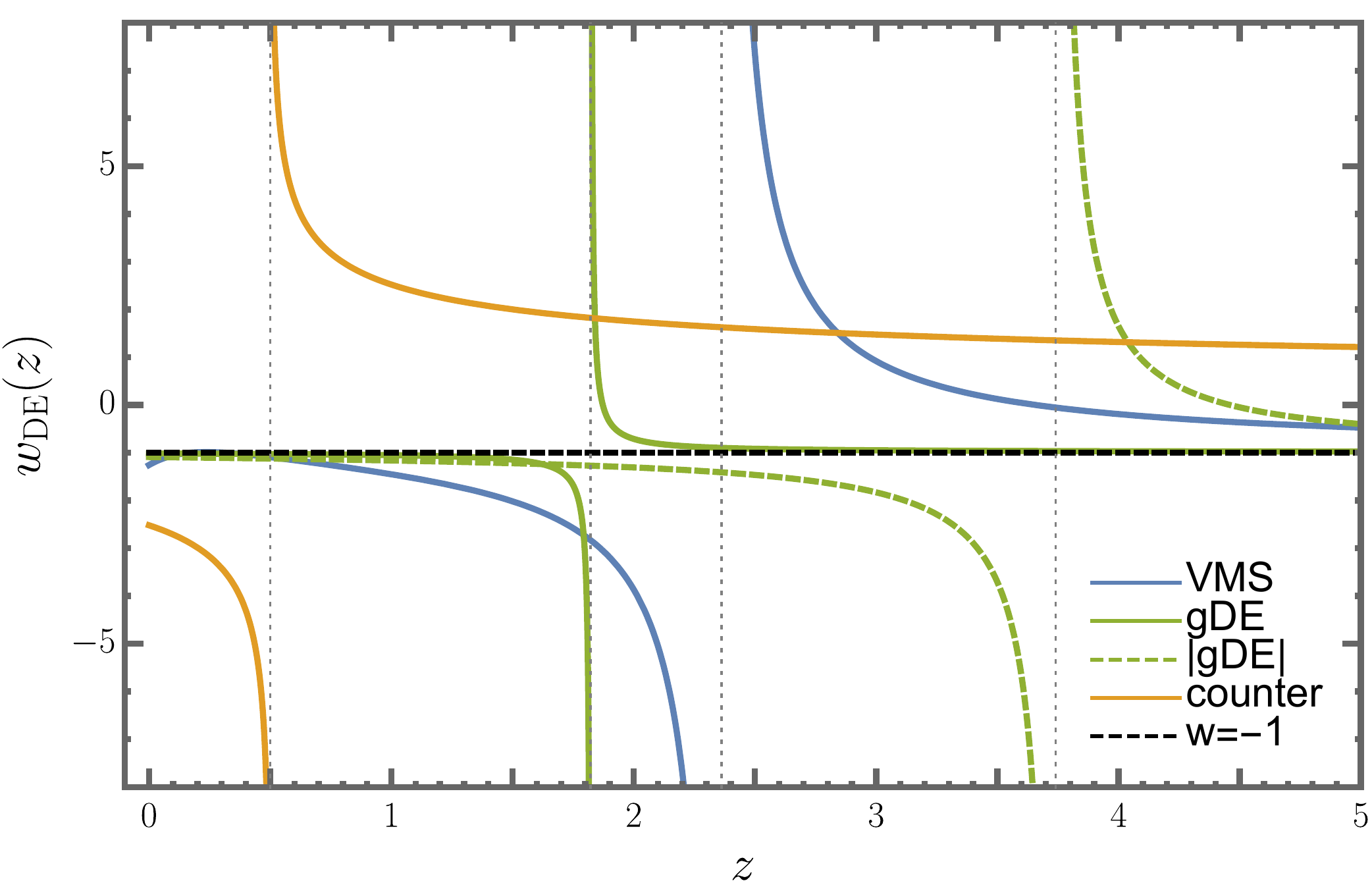}
    \caption{Top panel shows the DE densities for various models scaled by ${\rho_{\rm c,0}\equiv3H^2(z=0)}$ and the bottom panel shows their corresponding EoS parameters, both with respect to redshift $z$. The blue line corresponds to the mean values in Ref.~\cite{DiValentino:2020naf} from their analysis (CMB+all dataset) with $\alpha=1$ and the model is dubbed VMS for the initials of the authors. The solid green line corresponds to the gDE with mean values (for $\rho_{\rm DE,0}/\rho_{\rm c,0}$, $\lambda$, and $\gamma$) in the analysis of Ref.~\cite{Akarsu:2019hmw} when $\lambda$ is free. The dashed green line is again the gDE but when $(1-\lambda)$ is not a ratio of two odd integers (see~\cite{Akarsu:2019hmw} for details), the values used are $\rho_{\rm DE,0}/\rho_{\rm c,0}=0.6$, $\lambda=-1.36$, $\gamma=-0.09$. The yellow line is a counterexample to the converse statement for which the pole does not correspond to a zero of the density; it is characterized by ${\cal W}(z)=\frac{2}{(z-1/2)^{1/3}}$, and we used $\rho_{\rm DE,0}/\rho_{\rm c,0}=0.65$. }
    \label{fig:ex}
\end{figure}

The statement we will prove in this paper is that 
\cref{eq:cont}---whether or not it stems from general relativity---implies that the EoS parameter $w\equiv p/\rho$ of the fluid
has a pole at a point $z_p$ if $z_p$ is an isolated zero of $\rho(z)$ (i.e., if $\rho(z_p)=0$ and there exists a neighborhood of $z_p$ for which the only contained zero is $z_p$).
Moreover, this pole is such that (see Fig.~\ref{fig:ex})
\be
\lim_{z\to z_p^-}w(z)=-\infty,\,\,\text{and,}\,\,\lim_{z\to z_p^+}w(z)=\infty.\label{eq:singtype}
\ee
We only assume that $z_p$ is not an accumulation point for the poles of $w(z)$. 

The converse statement is not true: the EoS parameter might have a pole described exactly by \cref{eq:singtype}, yet the energy density does not necessarily vanish at $z_p$. We show this by a simple counterexample for which
${{\cal W}(z)=\frac{2}{(z-1/2)^{1/3}}}$ with $z_p=1/2$ (see the yellow lines in Fig.~\ref{fig:ex} and the definition of $\cal W$ in the relevant section).

\section{Dark energy as the source}
The above statement is clearly not limited to the DE density, however, for the rest of the paper we will consider the source whose energy density vanishes to be the DE. This is because the ambiguous nature of the DE lets us treat it as an effective source $\rho_{\rm DE}$ (e.g., a combination of a cosmological constant and a scalar field~\cite{Visinelli:2019qqu}, or just a modification to EFE due to an underlying modified gravity model such as in Ref.~\cite{Perez:2020cwa}) whose energy density can be negative and EoS parameter can attain values $w_{\rm DE}<-1$ and $w_{\rm DE}>1$ without necessarily violating energy conditions and having stability problems (see Ref.~\cite{Akarsu:2021fol} and references therein), and even be singular (these singularities are of the weak kind and spacetime is geodesically complete~\cite{Fernandez-Jambrina:2010ngm,Trivedi:2022ngt}). In addition, previous works showing how a negative DE density transiting to positive regime in the late universe is able to solve the prominent and mild cosmological tensions, emphasizes the importance of our results in the context of DE.

One might ask why should an effective DE obey the continuity equation. While this is not always necessary, it covers a broad family of cases. We can easily make the DE obey \cref{eq:cont}, for example, by writing
\begin{equation}
    G^{\mu\nu}=T^{\mu\nu}+T_{\rm DE}^{\mu\nu},
    \label{eq:EFE}
\end{equation}
for a RW universe, assuming DE to be a perfect fluid $T_{\rm DE}^{\mu\nu}=(\rho_{\rm DE}+p_{\rm DE}) u^\mu u^\nu+p_{\rm DE} g^{\mu\nu}$, and that rest of the sources do not interact with DE non-minimally; in this case, since  ${\nabla_\mu \qty(T^{\mu\nu}+T_{\rm DE}^{\mu\nu})=0}$ as a direct consequence of the twice contracted Bianchi identity, the minimal interaction condition implies 
\begin{equation}
    \nabla_\mu T_{\rm DE}^{\mu\nu}=0.\label{eq:contDE}
\end{equation}
Note that, for this particular example, if the DE density is to be negative at any given time, other physical sources or negative spatial curvature (spatially open universe) also must be present so that $H^2>0$ is satisfied.

\section{Proof}
In redshift form, when applied to the DE component, the continuity equation~\eqref{eq:cont} reads
\be
\dv{\rho_{\rm DE}(z)}{z} =  \frac{3}{1+z}\qty[1+w_{\rm DE}(z)]\rho_{\rm DE}(z),
\ee
where $p_{\rm DE}(z)=w_{\rm DE}(z) \rho_{\rm DE}(z)$. Then, rearranging and integrating the equation
\be
\int_{\rho_{\rm DE}(z_1)}^{\rho_{\rm DE}(z_2)}\frac{\dd{\rho_{\rm DE}'}}{\rho_{\rm DE}'}=3\int_{z_1}^{z_2}\dd{z}\frac{1+w_{\rm DE}(z)}{1+z}.
\ee
Defining 
\be
{\cal W}(z)\equiv3\frac{1+w_{\rm DE}(z)}{1+z},
\ee
and performing the integration on both sides of the above equation, we have
\be
\abs{\rho_{\rm DE}(z_2)}=\abs{\rho_{\rm DE}(z_1)}e^{\int_{z_1}^{z_2}{\cal W}(z)\dd{z}}.
\label{eq:rho}
\ee

Let $z_p$ be a zero of $\rho_{\rm DE}$, i.e., $\rho_{\rm DE}(z_p)=0$. We will show that $z_p$ is a pole of ${\cal W}(z)$.
First, observe that $z_p$ cannot be a zero of $\rho_{\rm DE}$ if ${\cal W}(z)$ is bounded from below in the neighborhood of $z_p$. Say, ${\cal W}(z)\geq r$ for an interval $\{z,z_p\}\in[x,y]$ around $z_p$ where $r$, $x$ and $y$ are arbitrary real numbers, and $z_p$ is the only zero of $\rho_{\rm DE}$ in that interval. The existence of such an interval without any other zero is guaranteed by our assumption that $z_p$ is an isolated zero of $\rho(z)$ (the isolation condition always holds by the identity theorem if $\rho(z)$ is an analytic function of redshift that does not vanish everywhere).
Then,
\begin{equation}
\begin{split}
    \int_{x}^{z_p}{\cal W}(z)\dd{z}\geq (z_p-x)r.
    \label{eq:bounded}
\end{split}
\end{equation}
From \cref{eq:rho} we have 
\be
\abs{\rho_{\rm DE}(z_p)}=\abs{\rho_{\rm DE}(x)}e^{\int_{x}^{z_p}{\cal W}(z)\dd{z}},
\ee
and from \cref{eq:bounded} we have
\be
\abs{\rho_{\rm DE}(x)}e^{\int_{x}^{z_p}{\cal W}(z)\dd{z}}\geq \abs{\rho_{\rm DE}(x)}e^{(z_p-x)r}>0.
\ee
Thus, ${\cal W}(z)$ cannot be bounded and there must be at least one pole of ${\cal W}(z)$ in the interval $[x,y]$, otherwise $\abs{\rho_{\rm DE}(z_p)}>0$ and $z_p$ is not a zero of $\rho_{\rm DE}$. Furthermore,
assuming that $z_p$ is not an accumulation point for the poles of ${\cal W}(z)$,
we can pick $[x,y]$ arbitrarily small around $z_p$ such that there exists only one pole in the interval $[x,y]$. This pole must be $z_p$ itself, because if it was not, we could have just picked a smaller $[x,y]$ interval around $z_p$ that exclude the single pole leaving no poles in the interval $[x,y]$, contradicting our initial statement that there must be at least one pole in the interval $[x,y]$ in order to satisfy $\rho_{\rm DE}(z_p)=0$.

Now, let us pick a small enough interval $[x,y]$ for which $z_p$ is the only pole of $\cal W$ and is the only zero of $\rho_{\rm DE}$ contained. Since $\rho_{\rm DE}(z)$ is continuous, from \cref{eq:rho}, we should have
\begin{equation}
\begin{aligned}
\abs{\rho_{\rm DE}(z_p)}&=\lim_{z_l\to z_p^-}\abs{\rho_{\rm DE}(x)}e^{\int_x^{z_l}{\cal W}(z)\dd{z}}\\&=\lim_{z_l\to z_p^+}\abs{\rho_{\rm DE}(y)}e^{-\int_{z_l}^{y}{\cal W}(z)\dd{z}},
\end{aligned}
\end{equation}
where the minus sign in the last exponent is because we switched the place of the integration bounds $y$ and $z_l$. The condition $\rho_{\rm DE}(z_p)=0$ is satisfied only for \begin{equation}
    \lim_{z_l\to z_p^-}\int_x^{z_l}{\cal W}(z)\dd{z}=-\lim_{z_l\to z_p^+}\int_{z_l}^{y}{\cal W}(z)\dd{z}=-\infty. \label{eq:last}
\end{equation}
Since we picked the interval $[x,y]$ such that ${\cal W}(z)$ is finite if $z\neq z_p$ for $z \in [x,y]$, Eqs.~\eqref{eq:last} are satisfied only for the appropriate signs of divergences as $z\to z_p$, i.e.,
\be
\lim_{z\to z_p^-}{\cal W}(z)=-\infty,\,\,\text{and,}\,\,\lim_{z\to z_p^+}{\cal W}(z)=\infty.
\label{eq:almost}
\ee
Clearly, \cref{eq:almost} is satisfied if and only if
\be
\lim_{z\to z_p^-}w_{\rm DE}(z)=-\infty,\,\,\text{and,}\,\,\lim_{z\to z_p^+}w_{\rm DE}(z)=\infty.
\ee
This concludes the proof.\footnote{While we have used redshift as the time parameter in our discussions and proof, everything holds for any other time parameter that can be mapped to redshift bijectively at least in the neighborhood of $z_p$ such as cosmic time $t$ in an ever expanding universe, or $y-$redshift~\cite{Cattoen:2007id} defined as $y\equiv\frac{z}{1+z}$; e.g., for the $y-$redshift, we have 
${\lim_{y\to y_p^-}w_{\rm DE}(y)=-\infty}$, and, $\lim_{y\to y_p^+}w_{\rm DE}(y)=\infty$,
where $y_p=\frac{z_p}{1+z_p}$. Note that the signs of the infinities or equivalently the directions of the one-sided limits should be interchanged if the alternative time parameter runs backwards to redshift as in the case of $t$.}

\section{Closing remarks}
The most important implication of this simple fact about the EoS is that one needs to be hesitant to use the EoS to characterize the DE in phenomenological statistical analyses. This point was emphasized also in Ref.~\cite{Wang:2018fng}. Even nonparametric reconstructions of the DE EoS parameter could fail to capture this singularity---an excellent example can be seen by comparing the results in Ref.~\cite{Escamilla:2021uoj} when $w_{\rm DE}(z)$ is constrained directly, and when it is inferred from a directly constrained $\rho_{\rm DE}(z)$. Since the present-day DE density is clearly positive, any nonsingular EoS parameter would result in a DE density that is always positive as it cannot cross zero without a pole; this unnecessarily restricts the phenomena of the DE density in observational analyses, since viewed as an effective source, the DE density may enjoy negative energy densities and singularities in its EoS parameter without repercussions. 

Another point is that, in the literature, an EoS parameter that satisfies $w<-1$ is dubbed ``phantom" and is strongly associated with an energy density that grows with the expansion of the universe. Although this is true for a phantom energy as defined in the seminal paper~\cite{Caldwell:1999ew} with a strictly positive energy density, the possibility of the energy density attaining negative values breaks this association. The more general interpretation that also covers cases that violates the positivity condition is that \textit{for $w>-1$, the energy density approaches to zero with the expansion, and for $w<-1$, it deviates from zero}. Of course, for a positive energy density, deviation from zero corresponds to growth as usual. See also the relevant discussion in Sec.~II of Ref.~\cite{Acquaviva:2021jov}.

It is worth noting that for a given present-day DE density value, the same EoS parameter can correspond to different histories for the DE density and the expansion. Take the blue line in the top panel of Fig.~\ref{fig:ex} as an example and make the transformation
\be
\rho_{\rm DE}(z\geq z_p)\to-\rho_{\rm DE}(z\geq z_p)
\ee
to see this. This transformation would leave the EoS parameter unchanged since the densities are present in \cref{eq:rho} with their absolute values. Note that, at least either one of the transformed or nontransformed energy densities produced by this method has to be nondifferentiable if $\dv{\rho_{\rm DE}}{z}\neq0$ at $z_p$.

Unlike the cases we investigated in this paper, if DE completely vanishes in the past, say, for $z\geq z_p$, the point of vanishing is still a zero of $\rho_{\rm DE}$, but it is not isolated. Thus, the above proof does not apply for this physically very interesting scenario; however, we state (the proof would be similar) that such a DE density would still have at least half of the pole, i.e.,
\be
\lim_{z\to z_p^-}w_{\rm DE}(z)=-\infty.
\ee
Note that such a form of the DE density that completely vanishes in the past can be smooth but it cannot be analytic.

Finally, we emphasize that, as discussed in the introduction, a negative DE density and the accompanied singularity/singularities in its EoS parameter are not problematic from the point of view of fundamental theories of physics and various theoretical models with such behavior already exists, especially if one considers an effective DE density rather than an actual physical source; moreover, a DE density that was negative in the past which has transited to the positive regime driving the present-day acceleration of the universe, is strongly motivated phenomenologically by cosmological observations. Future prospects related to this work could include checking whether the same or a similar behavior for the EoS parameter exists for anisotropic and/or inhomogeneous metrics for an energy density that touches zero, and specifically looking for an observational evidence for singularities in the DE EoS parameter.

\section{Acknowledgments}
The author thanks \"{O}zg\"{u}r Akarsu for valuable comments and acknowledges the support by The Scientific and Technological Research Council of Turkey (T\"{U}B\.{I}TAK) in scheme of 2211/A National PhD Scholarship Program.

\end{document}